# Consideration on relation between penetrated power and topological charge of millimeter-wave vortex in magnetized plasma

Chenxu Wang[1*], Hideki Kawaguchi[2], Hiroaki Nakamura[1,3], Shin Kubo[4]

[1]National Institute for Fusion Science, Toki, Japan 509-5292
[2]Muroran Institute of Technology, Muroran, Japan 050-8585
[3]Nagoya University, Nagoya, Japan 464-8601
[4]Chubu University, Kasugai, Japan 487-8501

*`naka-lab@nifs.ac.jp`

**Abstract.** It was demonstrated that the vortex field of a hybrid mode can propagate in the magnetized plasma region where plane waves are unable to propagate due to the cut-off condition. In this study, the dependence of the penetrated power of injected as millimeter-wave vortices of the hybrid mode in magnetized plasma is analyzed using the Finite-Difference Time-Domain (FDTD) method. The effect of the radius size of the corrugated waveguide on the penetrated power is described, revealing its significant contribution to reducing the deviation of topological charge in the hybrid mode. Furthermore, it was found that the penetrated power of the vortex field in magnetized plasma strongly depends on the topological charge $l$ and the deviation of topological charge.

**Keywords:** plasma-wave interaction, millimeter-wave vortex, hybrid mode, FDTD, magnetized plasma, waveguide.

## 1. Introduction

Optical vortices with a helical wavefront, which carry orbital angular momentum (OAM)[1] in addition to spin angular momentum (SAM), have been widely studied to explore the momentum exchange between light and matter. These studies have significant applications in various fields, such as optical trapping and manipulation of particles[2], optical communication systems utilizing multiplexing[3], and material processing through light-matter interaction[4-5]. Moreover, it was also shown that such phenomena of vortex fields exist in almost all frequency ranges, microwaves, millimeter waves, X-rays, $\gamma$-rays, ultralvilot and so on[6-9]. As a new application of optical vortices, for example, for vortex fields in millimeter wave frequency range, it was pointed out in 2021 by Tsujimura et al[10] that the millimeter-wave vortex assumed to be a Laguerre-Gaussian (L-G) mode can propagate in magnetized cold plasma even if the normal plane waves cannot propagate due to the cut-off condition. This means that more efficient plasma heating can be achieved by replacing the conventional plane wave with the vortex wave. In addition, it is numerically demonstrated that the millimeter-wave vortex of a hybrid mode[11] can propagate in uniform magnetized plasma where the normal plane wave cannot propagate due to the cut-off condition. It is also found that propagation power of the millimeter wave vortex in plasma strongly depends on the topological charge of the vortex[12]. In this paper, we discuss the better propagation conditions of millimeter-wave vortex of hybrid mode in uniform magnetized plasma by using full three-dimensional simulation of the finite-difference time-domain (FDTD) method.

## 2. Numerical formulation of hybrid mode in magnetized plasma

### 2.1. Millimeter-wave vortex of hybrid mode in corrugated waveguide

In this study, we assume that millimeter-wave vortex of the hybrid mode is generated in corrugated cylindrical waveguide and illuminated into magnetized plasma. It is known that there exists a stable propagation solution to the Helmholtz equation in a cylindrical corrugated waveguide in coordinates (*x, y, z*) and cylindrical coordinate $(\rho, \emptyset, z)$ exist as follows,

$$E_x = -iE_{mn}\frac{\beta}{k_c}J_{m-1}(k_c\rho) \times e^{i(\omega t - \beta z \pm (m-1)\emptyset)}, \tag{1-1}$$

$$E_y = -iE_{mn}\frac{\beta}{k_c}J_{m-1}(k_c\rho) \times e^{i\left(\omega t - \beta z \pm (m-1)\emptyset \pm \frac{\pi}{2}\right)}, \tag{1-2}$$

$$E_z = E_{mn}J_m(k_c\rho) \times e^{i(\omega t - \beta z \pm m\emptyset)}, \tag{1-3}$$

$$H_x = -iE_{mn}\frac{\beta}{Z_0 k_c}J_{m-1}(k_c\rho) \times e^{i\left(\omega t - \beta z \pm (m-1)\emptyset \pm \frac{\pi}{2}\right)}, \tag{1-4}$$

$$H_y = -iE_{mn}\frac{\beta}{Z_0 k_c}J_{m-1}(k_c\rho) \times e^{i(\omega t - \beta z \pm (m-1)\emptyset)}, \tag{1-5}$$

$$H_z = -iE_{mn}\frac{1}{Z_0}J_m(k_c\rho) \times e^{i(\omega t - \beta z \pm m\emptyset)}, \tag{1-6}$$

where $E_{mn}$ is amplitude of $E_z$ of *mn* mode, where *m, n* are integer. $J_m(x)$ is the first kind of Bessel's function of order of *m*, and $z_0 = \sqrt{(\mu_0/\varepsilon_0)}$ is wave impedance. It is assumed that the boundary conditions, $E_\emptyset(a, \emptyset, z) = 0$ is satisfied. The cut-off wave number $k_c = \rho_{mn}/a$ and inner radius *a* of corrugated cylindrical waveguide satisfy the following condition,

$$J_m(k_c a) = 0, \tag{2}$$

where $\rho_{mn}$ represents the *n*-th zero (*n*=1,2,3,…) of the Bessel's function $J_m(x)$ of the first kind of order *m* (*m*=0,1,2,3,…).

In particular, if we calculate the following Poynting power *P*,

$$P = \frac{1}{2T}\int_0^T dt \int_S (\boldsymbol{E} \times \boldsymbol{H}^*) \cdot d\boldsymbol{S}, \tag{3}$$

and the *z*-component of the orbital angular momentum $M_z$,

$$M_z = \frac{1}{2T}\int_0^T c\,dt \left( \int_S \left( \boldsymbol{\rho} \times \frac{1}{c^2}(\boldsymbol{E} \times \boldsymbol{H}^*) \right) \cdot d\boldsymbol{S} \right) \cdot \hat{z}, \tag{4}$$

for transverse cross-section *S* to propagation axis, the orbital angular momentum per unit power $M_z/P$ results in,

$$\frac{M_z}{P} = \frac{l}{\omega} = \frac{(m-1)+\sigma_z}{\omega}, \tag{5}$$

where $\sigma_z$=+1, -1 and 0 corresponds to left-hand, right-hand polarization and linear polarization. For example, if the millimeter-wave vortices are linearly polarized for the *x*-direction, which corresponds to $\sigma_z = 0$, the value of topological charge *l* of the vortex field of hybrid mode is (*m*-1). In particular, the solution (1) is plane wave $HE_{11}$ mode when *m*=1, which has no orbital angular momentum.

## 2.2. Drude-Lorentz macro model for magnetized plasma

We adopt the following Drude-Lorentz macro model as a model of magnetized plasma[12] by using electron displacement density vector ***P*** and current density vector $\boldsymbol{J} = d\boldsymbol{P}/dt$,

$$\frac{d\boldsymbol{J}}{dt} + \gamma \boldsymbol{J} + \omega_0^2 \boldsymbol{P} = \varepsilon_0 \omega_p^2 \left( \boldsymbol{E} + \frac{1}{n_e q_e}\boldsymbol{J} \times \boldsymbol{B}_0 \right), \tag{6}$$

where $\gamma$, $n_e$, $q_e$ and $\boldsymbol{B}_0$ are the dumping coefficient, the electron density, elementary charge and externally applied magnetic field, respectively. Then the FDTD formulation of the millimeter-wave vortex in magnetized plasma in 3D grid space for ***E***, ***H***, ***P*** and ***J*** is as follows,

$$\boldsymbol{E}^{n+1} = \frac{\frac{\varepsilon_0}{\Delta t} - \frac{\sigma}{2}}{\frac{\varepsilon_0}{\Delta t} + \frac{\sigma}{2}} \boldsymbol{E}^n + \frac{1}{\frac{\varepsilon_0}{\Delta t} + \frac{\sigma}{2}} \nabla \times \boldsymbol{H}^{n+\frac{1}{2}} - \frac{1}{\frac{\varepsilon_0}{\Delta t} + \frac{\sigma}{2}} \boldsymbol{J}^{n+\frac{1}{2}}, \tag{7}$$

$$\boldsymbol{H}^{n+\frac{1}{2}} = \boldsymbol{H}^{n-\frac{1}{2}} - \frac{\Delta t}{\mu_0} \nabla \times \boldsymbol{E}^n, \tag{8}$$

$$\boldsymbol{P}^{n+1} = \Delta t \boldsymbol{J}^{n+\frac{1}{2}} + \boldsymbol{P}^n, \tag{9}$$

$$\frac{\boldsymbol{J}^{n+\frac{1}{2}}-\boldsymbol{J}^{n-\frac{1}{2}}}{\Delta t} + \gamma \frac{\boldsymbol{J}^{n+\frac{1}{2}}+\boldsymbol{J}^{n-\frac{1}{2}}}{2} + \omega_0^2 \boldsymbol{P}^n = \varepsilon_0 \omega_p^2 \boldsymbol{E}^n + \frac{\varepsilon_0 \omega_p^2}{q_e n_e} \frac{\boldsymbol{J}^{n+\frac{1}{2}}+\boldsymbol{J}^{n-\frac{1}{2}}}{2} \times \boldsymbol{B}_0, \tag{10}$$

where $\Delta t$ is unit time step, and $\Delta t = 2.4975 \times 10^{-13}$ s. ***E*** and ***P*** are assigned to integer time step, ***H*** and ***J*** are assigned to half integer time step, that is, ***E*** and ***P*** or ***H*** and ***J*** are calculated

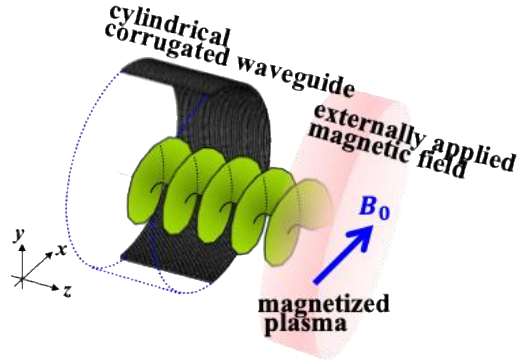

Figure 1: Overview of numerical model for plasma heating by hybrid mode vortex.

simultaneously.

## 3. Numerical examples

The overview of numerical model of the plasma heating by hybrid mode vortex is shown in Fig.1, it is assumed that the millimeter-wave vortex is excited inside of the corrugated cylindrical waveguide and travels to the downstream vacuum, and then illuminates the magnetized plasma with the uniform plasma density $n_e = 1.1 \times 10^{20}$ m$^{-3}$, where the externally applied magnetic field $|\boldsymbol{B}_0|$ is taken to be 2T, which is assumed to be oriented to the *x*-direction. The FDTD method numerical grid models (*y*-*z* plane) of millimeter-wave vortex in magnetized plasma are depicted in Fig.2. It is assumed that the millimeter-wave vortex of hybrid mode is excited at the two wavelengths distance from the upstream edge of corrugated waveguide, and then travels to the downstream vacuum and illuminates the plasma region. The Figs. 2(a)-2(c) are for three radius of waveguide *a*=30mm, 40mm and 60mm, respectively. In each figure, the numerical model in the *y-z* plane is shown in the upper, and the distribution of electric field intensity of incident vortex field in *x-y* cross section at vacuum is shown in lower. The whole grid size is taken to be $1000 \times 1000 \times 600$, and the unit grid size is 0.15mm, the frequency and power of the excited vortex field are assumed to be 77 GHz and 1MW.

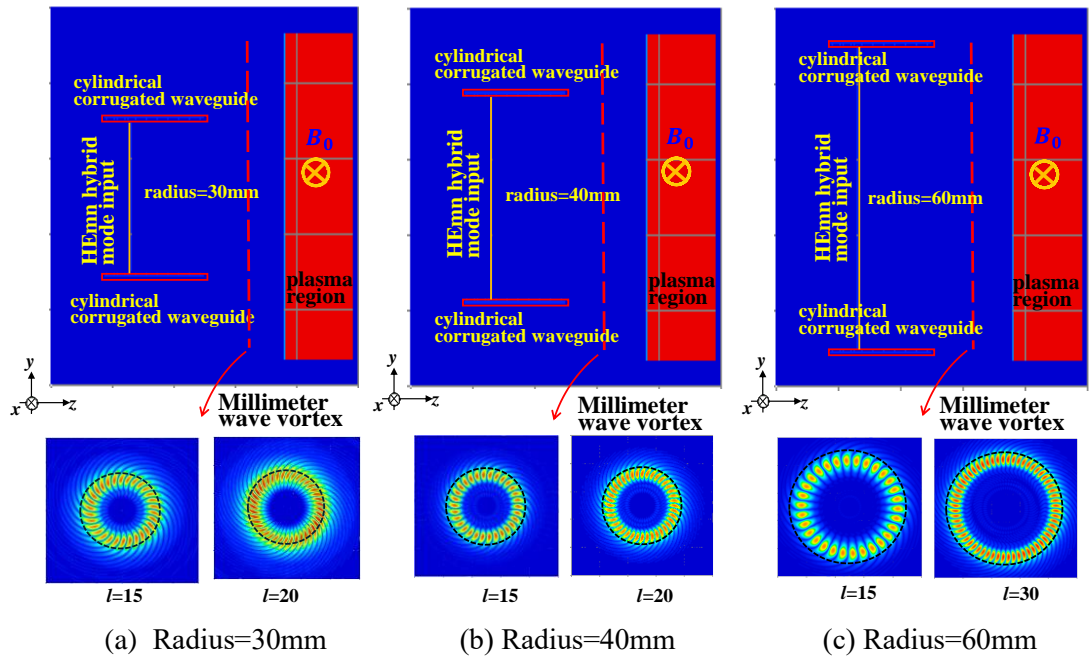

(a) Radius=30mm  (b) Radius=40mm  (c) Radius=60mm

Figure 2: FDTD simulation of numerical model

This study compares the penetrated power of millimeter-wave vortex in magnetized plasma and angular momentum of vortex fields for these of radius of waveguide *a*=30mm, 40mm and 60mm. In Fig. 3(a)-3(c), the penetrated power of millimeter-wave vortex in magnetized plasma for topological charge *l* is plotted in the upper, which is calculated by using (3) on a surface at the middle *z*-position inside the plasma region, and the lower table shows the comparison of theoretical topological charge *l* and numerically calculated angular momentum of $M_z/P \times \omega$, which is evaluated deviation of the topological charge by using (5). It can be seen from the simulation results of Fig.3 that the penetrated power of the millimeter-wave vortex in the magnetized plasma has peak value, and the peak value appears at the higher *l* as the radius of waveguide increases. In addition, it also can be seen from the lower tables that *l* and $M_z/P \times \omega$ has good agreements for lower *l*, and that the larger the radius of waveguide, the better the consistency between the input *l* and the calculated value $M_z/P \times \omega$. Accordingly, we can obtain

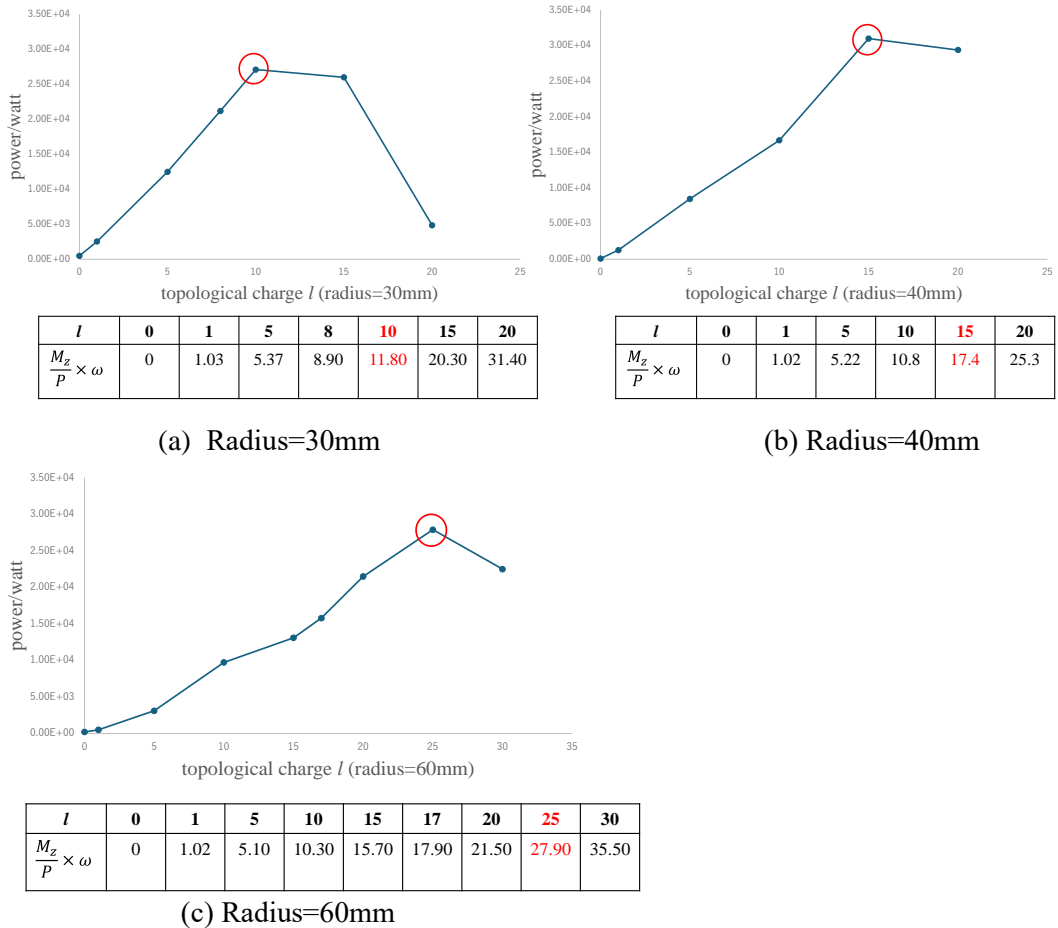

(a) Radius=30mm

| $l$ | 0 | 1 | 5 | 8 | **10** | 15 | 20 |
|---|---|---|---|---|---|---|---|
| $\frac{M_z}{P} \times \omega$ | 0 | 1.03 | 5.37 | 8.90 | **11.80** | 20.30 | 31.40 |

(b) Radius=40mm

| $l$ | 0 | 1 | 5 | 10 | **15** | 20 |
|---|---|---|---|---|---|---|
| $\frac{M_z}{P} \times \omega$ | 0 | 1.02 | 5.22 | 10.8 | **17.4** | 25.3 |

(c) Radius=60mm

| $l$ | 0 | 1 | 5 | 10 | 15 | 17 | 20 | **25** | 30 |
|---|---|---|---|---|---|---|---|---|---|
| $\frac{M_z}{P} \times \omega$ | 0 | 1.02 | 5.10 | 10.30 | 15.70 | 17.90 | 21.50 | **27.90** | 35.50 |

Figure 3: Propagation power of vortex fields in plasma for topological charge *l*

that when there is a good agreement between the input $l$ and numerically calculated value $M_z/P \times \omega$, which means lower deviation of the topological charge of the vortex, the penetrated power at the same radius of waveguide increases with the increase of topological charge.

## 4. Conclusion

In this study, the dependence of propagation of the hybrid mode millimeter-wave vortex in the magnetized plasma on topological charge $l$ has been discussed by using the FDTD method. The simulation results demonstrate that the penetrated power of millimeter-wave vortices in magnetized plasma is significantly influenced by the waveguide radius and the topological charge. In principle, penetrated power of millimeter-wave vortex to plasma increases for larger topological charge $l$. But efficiency of the power penetrated becomes worse when the deviation of topological charge of the vortex field is larger. In particular, waveguides with larger radius can propagate vortex fields with lower deviation of topological charge.

## Acknowledgment

The computation was performed using Research Center for Computational Science, Okazaki, Japan (Project: 24-IMS-C099) and Plasma Simulator of NIFS. The research was supported by KAKENHI (Nos. 21H04456, 22H05131, 23H04609, 22K18272, 23K03362), by the NINS program of Promoting Research by Networking among Institutions (01422301) by the NIFS Collaborative Research Programs (NIFS22KIIP003, NIFS24KIIT 009, NIFS24KIPT013, NIFS22KIGS002, NIFS22KISS021) and by the ExCELLS Special Collaboration Program of Exploratory Research Center on Life and Living Systems(24-S5).

## References

[1] ALLEN, Les, et al.: Orbital angular momentum of light and the transformation of Laguerre-Gaussian laser modes. Physical review A, 45:11 (1992), 8185.

[2] COJOC, Dan, et al.: Laser trapping and micro-manipulation using optical vortices. *Microelectronic Engineering*, 78(2005), 125-131.

[3] KHONINA, Svetlana Nikolaevna, et al.: Optical multiplexing techniques and their marriage for on-chip and optical fiber communication: a review. *Opto-Electronic Advances*, 5:8 (2022).

[4] OMATSU, Takashige, et al.: Metal microneedle fabrication using twisted light with spin. *Optics express*, 18:17 (2010): 17967-17973.

[5] TOYODA, Kohei, et al.: Using optical vortex to control the chirality of twisted metal nanostructures. *Nano letters*, 12:7 (2012), 3645-3649.

[6] BAHRDT, J., et al.: First observation of photons carrying orbital angular momentum in undulator radiation. *Physical Review Letters*, 111:3 (2013): 034801.

[7] Katoh, M., Fujimoto, M., Kawaguchi, et al.: Angular momentum of twisted radiation from an electron in spiral motion. *Physical Review Letters*, 118:9 (2017), 094801.

[8]  TAIRA Yoshitaka, KATOH Masahiro: Generation of optical vortices by nonlinear inverse Thomson scattering at arbitrary angle interactions. *The Astrophysical Journal*, 860:1 (2018), 45.

[9]  Taira, Yoshitaka, and Masahiro Katoh: Gamma-ray vortices emitted from nonlinear inverse Thomson scattering of a two-wavelength laser beam. *Physical Review A*, 98:5 (2018), 052130.

[10] Tsujimura, Toru Ii, and Shin Kubo: Propagation properties of electron cyclotron waves with helical wavefronts in magnetized plasma. *Physics of Plasmas*, 28:1 (2021).

[11] KAWAGUCHI Hideki, KUBO Shin: NAKAMURA Hiroaki, Orbital angular momentum of vortex fields in corrugated cylindrical waveguide hybrid mode. *IEEE Microwave and Wireless Technology Letters*, 33:2(2022), 118-121.

[12] Chenxu Wang, Hideki Kawaguchi, Hiroaki Nakamura, Shin Kubo: The study of propagation characteristics of millimeter-wave vortex in magnetized plasma by using FDTD method, *Jpn. J. Appl. Phys.*, 63:9 (2024), 09SP08.